\documentclass[10pt,a4paper]{article}
\usepackage{epsfig}
\usepackage{amsmath}
\usepackage{amssymb}
\usepackage{cite}
\usepackage{here}
\usepackage[scriptsize]{subfigure}
\usepackage[footnotesize,bf]{caption}

\newcommand{\reaktion}[1]{\mbox{$pp\rightarrow\,pp\:\!{#1}$}}

\newcommand{\dpheliumeta}{$d\vec{p}\rightarrow\, ^3$He\;\!$\eta$}
\newcommand{\pdheliumeta}{$\vec{p}d\rightarrow\, ^3$He\;\!$\eta$}

\begin{document}

\title{
\vspace*{-1.4cm}
{\large \bf Polarization observables in the $\eta$ meson production}\\
\vspace*{-.2cm}
{\small Letter of Intent for an experiment with a $\boldsymbol{4\pi}$ electromagnetic calorimeter @ COSY}\\
}

\author{
\normalsize {P.~Winter$^{1}$, R.~Czy\.zykiewicz$^{1,2}$, H.-H.~Adam$^{3}$, A.~Budzanowski$^{4}$,}\\
\normalsize {D.~Grzonka$^{1}$, M.~Janusz$^{2}$, L.~Jarczyk$^{2}$, B.~Kamys$^{2}$, A.~Khoukaz$^{3}$,}\\
\normalsize {K.~Kilian$^{1}$, P.~Moskal$^{2}$, W.~Oelert$^{1}$, C.~Piskor-Ignatowicz$^{2}$, J.~Przerwa$^{2}$,}\\
\normalsize {T.~Ro\.zek$^{1,5}$, T.~Sefzick$^{1}$, M.~Siemaszko$^{5}$, J.~Smyrski$^{2}$, A.~T\"aschner$^{3}$,}\\
\normalsize {M.~Wolke$^{1,6}$, P.~W\"ustner$^{7}$, W.~Zipper$^{5}$}\\
\normalsize {(COSY-11 Collaboration)}\\
}

\date{}
\maketitle

\begin{center}

\vspace*{-.6cm}
\begin{small}
{\em $^{1}$ IKP, Forschungszentrum J\"ulich, D-52425 J\"ulich, Germany}\\
{\em $^{2}$ Institute of Physics, Jagellonian University, PL-30-059 Cracow, Poland}\\
{\em $^{3}$ IKP, Westf\"alische Wilhelms--Universit\"{a}t, D-48149 M\"unster, Germany}\\
{\em $^{4}$ Institute of Nuclear Physics, PL-31-342 Cracow, Poland}\\
{\em $^{5}$ Institute of Physics, University of Silesia, PL-40-007 Katowice, Poland}\\
{\em $^{6}$ Svedberg Laboratory, Thumbergsv\r{a}gen 5A, Box 533, S-75121 Uppsala, Sweden}\\
{\em $^{7}$ ZEL, Forschungszentrum J\"ulich, D-52425 J\"ulich, Germany}\\
\end{small}
\end{center}
\vspace*{.4cm}

\begin{abstract}
We propose to extend the investigations of the $\eta$ meson production in nucleon-nucleon collisions by studying single and double polarization observables. Over the last years, the $\eta$ meson has been studied at several different experiments mainly by measurements of unpolarized total and differential cross sections. To reveal the production mechanism, polarization observables help to disentangle the amplitudes of the partial waves and therefore to understand the underlying exchange processes. Analyzing powers have been measured enabling to separate the contributing amount of some of  these amplitudes. Double polarization observables will widen the number of accessible amplitudes and hence give further constraints to the theoretical models. Due to the trigonometric symmetries in polarization measurements, a symmetric detector geometry in the azimuthal angle is preferable. The WASA detector with its $4\pi$ geometry for neutral and charged particles together with the use of a frozen spin target should therefore allow to study double polarization observables not only in $\vec{p}\vec{p}$ but also in $\vec{p}\vec{d}$ or $\vec{d}\vec{p}$ collisions. This first requires the development of a frozen-spin target whereas single polarization observables can be measured directly with the current setup.
\end{abstract}

\newpage

\section{Physics case}
\subsection{Scientific justification}
\begin{figure}[ht]
\begin{center}
\subfigure[\label{pp-invmass}]{\epsfig{file=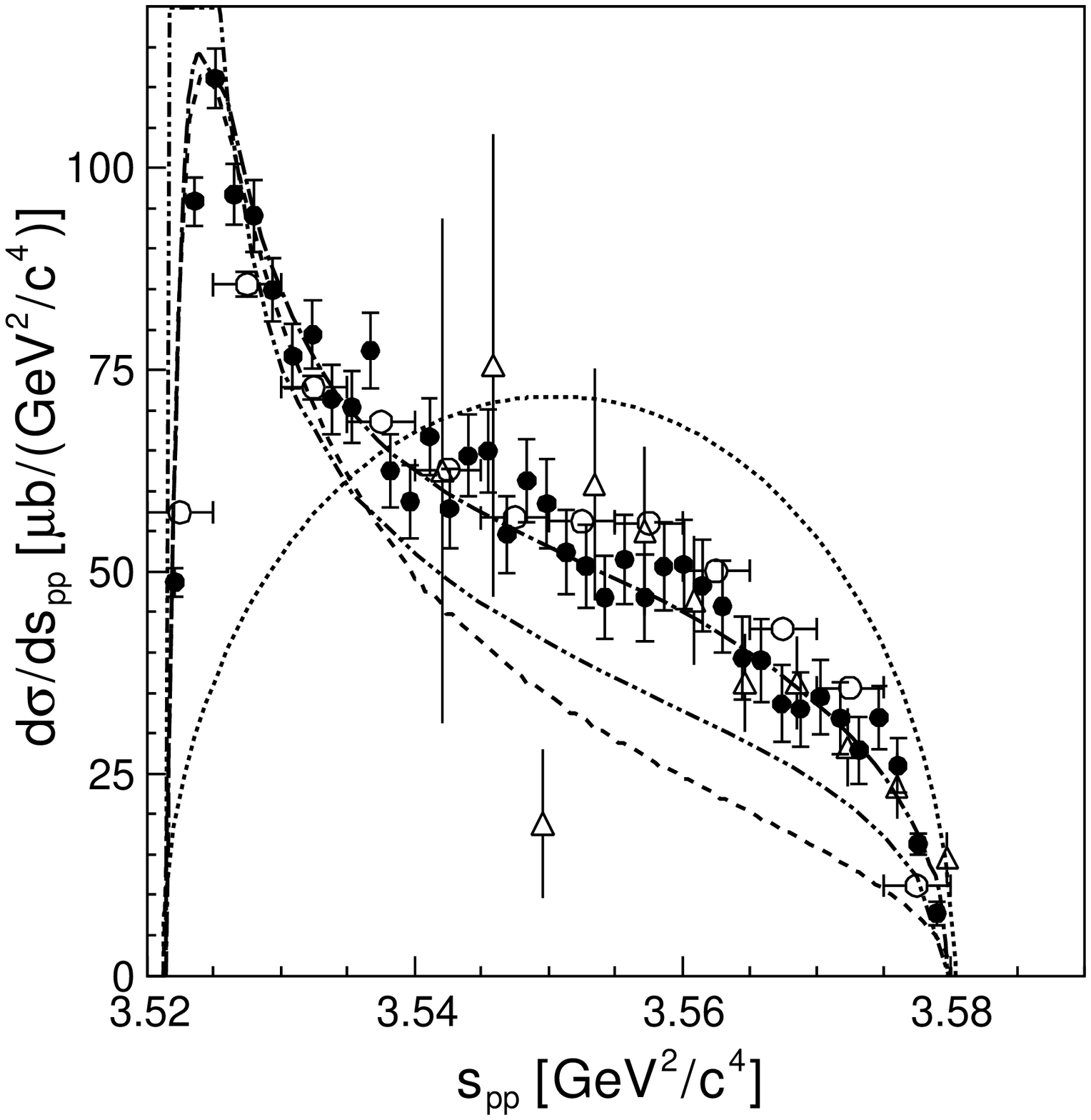,scale=0.35}}
\hspace{0.5cm}
\subfigure[\label{energydep}]{\epsfig{file=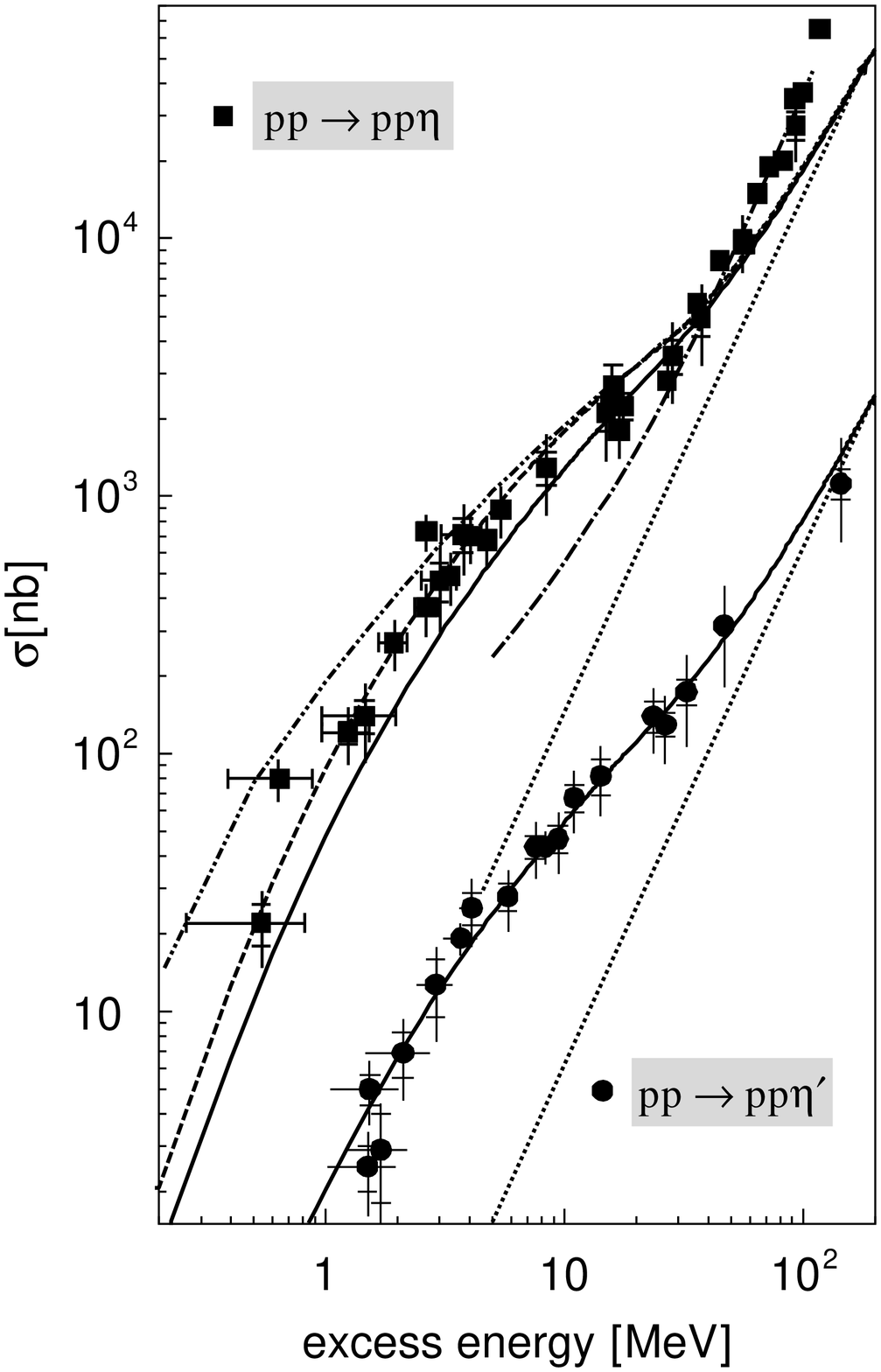,scale=0.25}}
\end{center}
\caption{a) Squared $pp$ invariant mass in the reaction \reaktion{\eta} (solid circles: \mbox{COSY-11} at $Q=15.5$\,MeV \cite{moskal:03-2}, open circles: COSY-TOF at $Q=15$\,MeV \cite{abdelbary:02}, open triangles: PROMICE/WASA at $Q=16$\,MeV \cite{calen:99}). The dotted and dashed lines correspond to the phase space and its modification by an incoherent pairwise treatment of the three-body FSI, respectively. The inclusion of a P-wave \cite{nakayama:03-1} is depicted by the dashed-dotted line and a full three-body treatment with a pure s-wave \cite{fix:02} by the dashed-double-dotted line. 
b) Energy dependence of the total cross section for the reactions \reaktion{\eta} (squares \cite{bergdolt:93, chiavassa:94, calen:96, calen:97, hibou:98, smyrski:00}) and \reaktion{\eta'} (circles \cite{hibou:98, moskal:98-2, moskal:00-2, balestra:00-2, khoukaz:04}). The lines are corresponding to those of the figure \ref{pp-invmass}. The solid line is the phase space modified by the $pp$ FSI including the Coulomb interaction.}
\end{figure}

The $\eta$ meson was intensively studied in nucleon nucleon collisions at several different experiments worldwide. Total cross sections \cite{bergdolt:93, chiavassa:94, calen:96, calen:97, hibou:98, smyrski:00}, as well as their differential distributions \cite{calen:99, tatischeff:00, moskal:03-2, abdelbary:02} have been the main subject of these investigations. Three independent experiments \cite{calen:99, abdelbary:02, moskal:03-2} recently revealed an unexpected enhancement in the proton-proton invariant mass spectrum m$_{pp}$ at higher $pp$ invariant mass. One theoretical explanation reproduces its shape by introducing higher partial waves \cite{nakayama:03-1}. Although the inclusion of the $^1$S$_0\rightarrow\,^3$P$_0$s transition\footnote{For details on the notation of the partial waves see for example \cite{meyer:01}.} in addition to the dominant $^3$P$_0\rightarrow\,^1$S$_0$s threshold amplitude describes the $pp$ invariant mass distribution very well (dashed-dotted line in figure \ref{pp-invmass}), this approach fails in reproducing the total cross section at excess energies below $Q=40\,$MeV (dash-dotted line in figure \ref{energydep}) by significantly underestimating the experimental cross section. The inclusion of the $\eta$N final state interaction (FSI) might help to resolve this issue.\\

In comparison, a rigorous full three-body treatment with only $s$-wave in the final state \cite{fix:02} decreases the differencial cross section at larger $pp$ invariant mass, while the total cross section is enhanced at low $Q$-values compared to the data (see the dashed-double-dotted lines in figures \ref{pp-invmass} and \ref{energydep}). The neglect of the Coulomb interaction should account partly of this latter discrepancy. The effects of the full three-body treatment on the total cross section as well as the $pp$ mass distribution are thus opposite compared to a P-wave admixture in the $pp$ subsystem. Therefore, this calculation does not reproduce the invariant mass spectrum.

A recent exploration with a more phenomenological analysis \cite{deloff:03} treating the $NN$-FSI effects rigorously reveals that the inclusion of a weak energy dependence of the dominant $^1$S$_0$ amplitude rather than higher partial waves reproduces the $pp$ mass distribution for $Q=15.5$\,MeV very well\footnote{The corresponding lines were not included to the figures \ref{pp-invmass} and \ref{energydep} but can be reviewed in the reference \cite{deloff:03}.}. The description also reproduces quite well both the m$_{\eta p}$ mass distribution and the energy dependence of the total cross section. For the data at $Q=41$\,MeV, the energy dependence of the $^1$S$_0$ amplitude alone fails to describe the experimental data.

\begin{figure}[ht]
\begin{center}
\epsfig{file=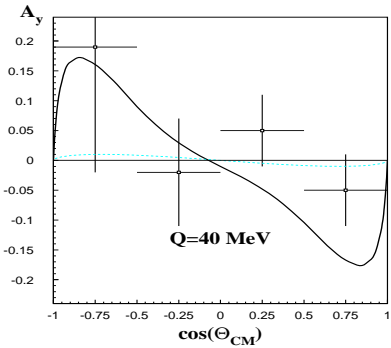,width=0.5\textwidth,height=0.4\textwidth}
\end{center}
\caption{Analyzing power as a function of the $\eta$ polar angle in the center of mass. Data are from  \cite{winter:02-2} while the curves are one-boson exchange models (solid line \cite{nakayama:03-1}, dashed line \cite{faeldt:01}).\label{analyzing}}
\end{figure}

First steps in direction of a more complete database on the $\eta$ meson production were performed \cite{winter:02-2,winter:02-3,roderburg:02,czyzyk:02} with polarization observables. They give further insight into the understanding of the production mechanism by a decomposition of the contributing partial waves. Such a procedure was used in its full complexity already in the $\pi^0$ production \cite{meyer:01} in $\vec{p}\vec{p}\rightarrow pp\pi^0$. Here, the double polarized measurements in principle allow for a complete extraction of partial waves up to D waves in the $pp$ system. A similar but less complex procedure in case of the $\eta$ meson has been applied for the single polarized data at $Q=40$\,MeV \cite{winter:02-2} in order to extract partly the accessible partial wave amplitudes. Up to now both the low statistics and a beam polarization of $0.5\pm 0.1$ limited, however, the interpretation of these results. Figure \ref{analyzing} shows the data obtained at the COSY-11 experiment \cite{winter:02-2} in combination with two scenarios \cite{nakayama:03-1,faeldt:01} for the excitation of the S$_{11}$(1535) resonance which plays a dominant role in the $\eta$ production according to the general opinion. Both theoretical approaches are based on one-boson exchange. The solid line in figure \ref{analyzing} represents the results from \cite{nakayama:03-1} in which the excitation process is mainly driven by $\pi$ and $\eta$, whereas the dashed line represents a dominant $\rho$ exchange \cite{faeldt:01}.

The unclear theoretical interpretation of the high statistics data on the $\eta$ production and the not yet understood production mechanism give the necessity for more detailed studies. Polarized measurements will be a key to verify the different scenarios mentioned above as well as to judge on the details of the excitation of the S$_{11}$ resonance. Here, the yet unmeasured double polarization observables, such as for instance the spin correlation functions, become important. As it was shown in the reference~\cite{nakayama:03-1} measurements of the spin correlation functions (either C$_{xx}$ or C$_{yy}$) will allow to infer about the contributions of the final $^{3}$P$_{0}$s and $^{3}$P$_{2}$s partial waves, which information is inaccessible from the single polarization observables in a model independent way. This observable will also provide the information (model independent) about the $pp$ S-wave contribution in the final state~\cite{nakayama:03-1}. 

In combination with a polarized deuteron beam or target, the investigations on the $\eta$ could be continued by exploring the reaction \pdheliumeta, closely related with the question concerning a $^3$He\;\!$\eta$ bound state. 

\subsection{Existing data}
Naturally, the first obtained observables concerning the close-to-threshold production of the $\eta$ meson  were the total cross sections for the $pp\to pp\eta$ reaction measured over a wide range of the excess energies~\cite{bergdolt:93, chiavassa:94, calen:96, calen:97, hibou:98, smyrski:00} (see figure~\ref{energydep}). Existing data gathered worldwide at different facilities are in a good agreement between each other and as it was shown in \cite{moskal:02-3} the close-to-threshold behaviour of the total cross section can be well reproduced by the phase-space and both pp and p$\eta$ final state interactions (FSI). Neglecting the p$\eta$ FSI leads to the underestimation of the total cross section in the excess energy range between 0 and 40 MeV. 

These fundamental measurements were followed by the investigations on the angular distributions for the $pp\to pp\eta$ reaction \cite{calen:99, tatischeff:00, moskal:03-2, abdelbary:02} as well as the final pp and p$\eta$ invariant mass distributions~\cite{moskal:03-2, abdelbary:02}. As the observed $\eta$ angular distribution in the overall center-of-mass frame is consistent with an isotropic emission~\cite{moskal:03-2, abdelbary:02} it appears that the $\eta$ meson is predominantly produced in the $s$-wave.

On the quasi-free sector, measurements of the total cross section in the $pn\to pn\eta$ reaction were performed~\cite{calen:98, moskal:03-3}, where the neutron originates from a deuteron target. The investigations performed by the WASA/PROMICE collaboration yielded a ratio R$_{\eta} = \frac{\sigma(pn\to pn\eta)}{\sigma(pp\to pp\eta} = 6.5$ in the excess energy range between 16 MeV and 109 MeV, demonstrating a strong isospin dependence of the production process of the $\eta$ meson. The analysis of the COSY-11 data concerning this reaction is still in progress.

Measurements of the analyzing power A$_y$ for the $\vec{p}p\to pp\eta$ reaction performed by the COSY-11 collaboration~\cite{winter:02-2} show rather small values of A$_y$. Within the relatively large error bars the values are practically equal to zero, which again suggests that the production of the $\eta$ meson takes place in the $s$-wave solely. However, in order to confirm this hypothesis a higher accuracy is required. Results of the analysis of the data taken by the COSY-11 group at $Q=10$ and $Q=40$\,MeV will be provided soon, while the statistical uncertainities will be reduced by a factor of about 2.

\subsection{Why a $\boldsymbol{4\pi}$ detector?}
Within current COSY experiments, in case of the \reaktion{\eta} reaction the 4-momenta of the two outgoing protons are measured and with the known beam momentum, the 4-momentum of the $\eta$ meson is determined. The high precision momentum measurement in combination with the precise COSY beam results in well defined $\eta$ peaks but the missing mass technique is connected with an unavoidable physical background. As far as total or differential cross sections like angular distributions are concerned the separation of the $\eta$ events from the background is no problem but the extraction of the full, for the 3-body system, five-fold differential distribution or even partly integrated distributions like Dalitz plots is impossible. These highly differential distributions are necessary for a detailed analysis. Also with regard to polarization observables, the detection of all charged and neutral particles is therefore necessary in future experiments. In addition, polarized measurements favor a symmetric detector setup due to the trigonometric symmetries in polarization observables. The 4$\pi$ WASA detector fulfills already in the current setup these requirements, where the $\eta$ meson can be clearly identified via its decay into 2$\gamma$ (see for details \cite{calen:99} and references therein).  

In comparison, current experiments at COSY are either limited by acceptance or resolution for the extension of the $\eta$ meson production studies. Due to the unavoidable physical background in the missing mass technique, the necessity for a symmetric $4\pi$ detector for charged and neutral particles became obvious for a better understanding of the $\eta$ meson production, that is single and double polarized measurements.\\

\section{Experimental details}
\newcommand{\extdensity}{$10^22/$cm$^$}
\newcommand{\intdensity}{$10^14/$cm$^$}
The $\vec{p}\vec{p}\to pp\eta$ reaction will be identified via the measurement of a four-fold coincidence of two protons together with $2\gamma$ in the final state. Besides that one can also detect other decay channels whereas these branches will finally lead to $\ge 3\gamma$ and therefore their experimental treatment remains similar. However, it might turn out in more detailed studies that these channels could be better suited with regard to the signal-to-background ratio and resolution. In the following, we therefore want to illustrate the experimental issues on the $2\gamma$ decay with a branching ratio $B=39.43\pm 0.26\%$~\cite{hagiwara:02} which has to be taken into account in the further determination of the total production rate.

\begin{figure}[ht]
\begin{center}
\epsfig{file=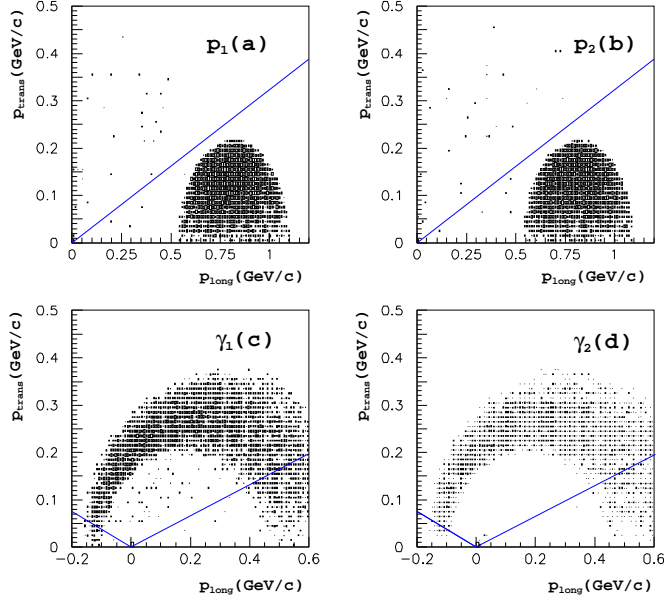,scale=0.45}
\end{center}
\caption{Simulated transverse versus longitudinal momenta for the reaction \reaktion{\eta}$\to p^{(a)}p^{(b)}\gamma^{(c)}\gamma^{(d)}$ with the particles' superscripts corresponding to the according panels. The photons' acceptance in the CsI calorimeter is indicated by the straight lines. In case of the protons the line indicates the separation between the forward and the central part of the detector. \label{ppeta-mc}}
\end{figure}

Furthermore, an overall detector efficiency $\mathcal{E}$ for the setup has to be included in the calculations. Since in this reaction, we intent to study the polarization observables up to a minimum of $Q=40\,$MeV\footnote{This value should not be regarded as a fixed upper limit, since higher momenta in COSY are accessible without any problems. The same holds for the $^3$He$\;\!\eta$ production.}, the required beam momenta for the measurements will be $1.982-2.100\,$GeV/c. Thus, we restrict the following considerations to 2 representative $Q$-values. For an excess energy $Q=40$\,MeV the overall efficiency for the $pp$ detection is about 90$\%$ and for the $2\gamma$ $\approx66\%$ \cite{wolke:03}. Although for $Q=10$\,MeV the corresponding values are not known, they should be roughly of the same order. A qualitative illustration of these numbers\footnote{Not quantitative, since the pictures only show the geometrical acceptance and not the efficiency.} is presented in figure \ref{ppeta-mc} where simulated transverse versus longitudinal momentum distibutions of the final state particles are shown. The simulations were performed assuming a phase space distribution weighted with the $pp$ FSI. The obviously small fraction of lost photons in the forward and backward cones underlines the reasonability of the above mentioned efficiency. 

For technical reasons, an internal  polarized storage cell target will be not feasible \cite{rathmann:03}. An external frozen spin target \cite{bradtke:99} is momentarily limited to 10$^6/$s beam particles. Therefore, with a reasonable target thickness of $6\cdot10^{23}$cm$^{-2}$ one can at present assume a luminosity of $\mathcal{L} = 6\cdot 10^{29}$\,cm$^{-2}$s$^{-1}$. Altogether with the total cross sections of $\sigma(Q=10\,$MeV) = 1.5$\,\mu$b and $\sigma$($Q=40\,$MeV) = 6\,$\mu$b in case of $\vec{p}\vec{p}\to pp\eta$, we obtain for the number of events per day:
\begin{itemize}
\item
$Q=10$\,MeV:
\begin{eqnarray*} 
N &=& \mathcal{L} \cdot B \cdot \mathcal{E} \cdot \sigma \cdot t\\
  &=& 6\cdot 10^{29}\mbox{cm$^{-2}$s$^{-1}$} \times 0.3943 \times 0.6 \times 1.5\cdot 10^{-30}\mbox{cm$^2$} \times 86400\,\mbox{s}\\
&\approx& \underline{\underline{18\,000}}
\end{eqnarray*}
\item
$Q=40$\,MeV:
\begin{eqnarray*} 
N &=& 6\cdot 10^{29}\mbox{cm$^{-2}$s$^{-1}$} \times 0.3943 \times 0.6 \times 6\cdot 10^{-30}\mbox{cm$^2$} \times 86400\,\mbox{s}\\
&\approx& \underline{\underline{73\,000}}
\end{eqnarray*}
\end{itemize}

\begin{figure}[ht]
\begin{center}
\epsfig{file=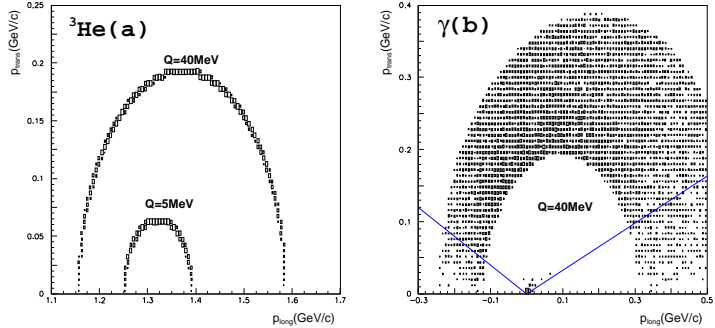,scale=0.48}
\end{center}
\caption{Simulated transverse versus longitudinal momenta of the outgoing helium in the reaction \pdheliumeta $\to\ ^3$He$^{(a)}\:\!\gamma^{(b)}\gamma$. The straight lines indicate the photon acceptance of the CsI calorimeter. \label{pdhelium}}
\end{figure}

For the identification of the reaction\footnote{In this case, it should be left an open issue for now, whether the equivalent reaction \dpheliumeta\ is better suited for reasons of a higher luminosity and/or better acceptance.} \pdheliumeta, a three-fold coincidence between $^{3}$He and 2$\gamma$ will be required. Here, the necessary beam momenta for a scan of the energy dependence of the appropriate polarization observables ranges from p$_{beam}=1572$ to $1654$\,MeV/c. In comparison to the $pp\eta$ case, the main difference should occur from the lower cross section (roughly a factor of 10). At an external location at COSY, there is in principle no problem for the detection of the $^3$He in the forward direction with a good acceptance, e.g. in combination with the COSY-TOF as the forward detector. Therefore, we again show the simulated momenta for the outgoing particles in figure \ref{pdhelium} but we do not repeat the calculation of the number of events which will be about ten times lower. Here, all $^3$He up to at least $Q=40$\,MeV lie within the forward detector area.

Concluding, double polarization observables will be accessible for the two final states ($pp\;\!\eta$ and $^3$He\;\!$\eta$\footnote{Therefore, also polarized deuterons are necessary which is experimentally no problem. But in that case, further theoretical work has to be done to correct for the non negligible d-wave contribution in the deuteron wave function.}) at an external location for the WASA detector setup at COSY in combination with a frozen spin target. 
The polarization programme on the $\eta$ production will be started with the single polarization observable prolonging the time for the development of a polarized target. \\

We would like to thank Christoph Hanhart for his advices on the theory and Frank Rathmann for the comments on polarized targets.

\bibliography{abbrev,polarisation,general}

\end{document}